\documentclass[twoside]{ilcws10}
\usepackage[latin1]{inputenc}
\usepackage[dvips]{graphicx,epsfig,color}
\usepackage{wrapfig,rotating}
\usepackage{amssymb,amsmath,array}

\begin{document}
\title{top FCNC physics at a Linear Collider after the LHC
} 
\author{Renato Guedes$^1$, Rui Santos$^2$ and Miguel Won$^{2,3}$
\thanks{RG is supported by FCT, grant SFRH/BPD/47348/2008.
RS is supported by the FP7 via a Marie Curie IEF,  PIEF-GA-2008-221707. MW is supported by FCT, grant SFRH/BD/45041/2008.}
\vspace{.3cm}\\
1- Centro de F\'{\i}sica Te\'orica e Computacional, Faculdade
de Ci\^encias, Universidade de Lisboa,\\
Avenida Professor Gama Pinto, 2, 1649-003 Lisboa, Portugal; \\
\vspace{.1cm}\\
2- NExT Institute and School of Physics and Astronomy,\\
        University of Southampton, Highfield, Southampton SO17 1BJ, UK; \\
\vspace{.1cm}\\
3- LIP- Departamento de F\'{\i}sica, Universidade de Coimbra\\
3004-516 Coimbra, Portugal.\\
}

\maketitle

\begin{abstract}
Flavour changing neutral currents are absent at tree level in the Standard Model (SM) and are highly suppressed at higher orders due the Glashow-Iliopoulos-Maiani mechanism. Hence, any evidence of such currents related to the top quark will definitely signal new physics beyond the SM. In this work we will discuss how an electron-positron collider can contribute to the field after the Large Hadron Collider (LHC) has collected at least 100 $fb^{-1}$ of integrated luminosity operating at $\sqrt{s}=14 \, TeV$.
\end{abstract}
\section{Why top FCNC?}
\label{sec:why}

The top is the least known of all quarks. In the SM the top decays almost exclusively to a $b$-quark and a $W$ boson. Furthermore, top decays via flavour changing neutral currents (FCNC) are absent at tree level in the SM and the loop-induced decays are severely constrained due the Glashow-Iliopoulos-Maiani (GIM) mechanism. The largest value for any top FCNC branching ratio in the SM is $ {\cal O} (10^{-12})$ for the $t \to c g$ decay~\cite{Eilam:1990zc, AguilarSaavedra:2004wm}, which is clearly out of reach of present and future colliders. Hence, any hint of FCNC related to the top quark, however small, would definitely signal new physics related to flavour. A number of different models discussed in the literature predict a huge enhancement of the top FCNC branching ratios: from quark-singlet models~\cite{AguilarSaavedra:2004wm, AguilarSaavedra:2002ns} to two-Higgs doublet models~\cite{Atwood:1996vj}, from Supersymmetry (see ~\cite{AguilarSaavedra:2004wm} for a discussion)  to Technicolour~\cite{He:1998ie}, branching ratios of the order of $ {\cal O} (10^{-4})$ or $ {\cal O} (10^{-5})$ can be reached in some best-case scenarios. Therefore, some of these theories could actually be probed at the LHC. In this work we will address the following question: will there still be top FCNC physics to explore after the LHC at an electron-positron and/or at a photon collider? Such a study would require a precise knowledge about the total luminosity that will be collected at the LHC. Moreover, it is still not clear when a new electron-positron or photon-photon collider will be built let alone its centre-of-mass energy and luminosity. Hence, we have chosen as the "future" the LHC at $100 fb^{-1}$ and have relied on the benchmarks available in the literature for proposed electron-positron and $\gamma \gamma$ colliders. Evidently, the luminosity collected at the LHC could reach $300  fb^{-1}$ or more (with the Super Large Hadron Collider) before any other collider starts operation. However, it is not clear what the effect on bounds obtained for top FCNC related observables would be and it is reasonable to assume that most of limits will stay within the same order of magnitude due to the difficulties of operating at very high luminosity. We will focus on the scenario where no evidence for new physics is found - otherwise a different approach has to be taken to understand what is the vertex or vertices that give the main contribution to the new physics observed.

In the next section we will first set the framework for our study; we will then proceed to a review of the results on top FCNC physics and the predictions available for the LHC; finally we will discuss top FCNC physics at a linear collider and at a photon collider.
\section{A framework for top FCNC - the effective operator approach}
\label{sec:frame}

All the calculations presented in this work will be done using the effective operator approach. This formalism~\cite{Buchmuller:1985jz} is based on the assumption that new physics beyond the SM will reflect itself at low energies as operators of dimension larger than four. The building blocks of this new theory, which has the SM as its low energy limit, are the SM fields (although other fields could be added) and will be assembled in accordance with the SM gauge symmetries $SU(3)_C \times SU(2)_W \times U(1)_Y$.
We write the lagrangian as
\begin{equation}
{\cal L} \;\;=\;\; {\cal L}^{SM} \;+\; \frac{1}{\Lambda}\,{\cal L}^{(5)} \;+\; \frac{1}{\Lambda^2}\,{\cal L}^{(6)} \;+\; O\,\left(\frac{1}{\Lambda^3}\right) \;\;\; , \label{eq:l}
\end{equation}
where ${\cal L}^{SM}$ is the usual SM lagrangian, ${\cal L}^{(5)}$ and ${\cal L}^{(6)}$ are the dimension five and six lagrangians and $\Lambda$ is the energy scale for which one expects physics beyond the SM to become relevant. ${\cal L}^{(5)}$ is discarded since it breaks baryon and lepton number.

The number of dimension six operators that obey the SM symmetries is huge and was presented in~\cite{Buchmuller:1985jz} where several artifices from Fierz transformations to the equations of motion were used to reduce the number of operators to a minimum of independent operators. However, the operators in~\cite{Buchmuller:1985jz} were defined up to flavour indices and a further reduction in the number of operators can be achieved depending on the physical process under study. We are considering operators that have at least one top quark and where FCNC occurs. These operators can have their origin in the strong sector or in the electroweak sector or can be written as four-fermion operators. A discussion on the operators we have used for the strong sector can be seen in~\cite{Ferreira:2005dr} while for the electroweak sector the operators are defined in \cite{Han:1998yr, Grzadkowski:2003tf, Ferreira:2008cj, Coimbra:2008qp}. Four-fermion operators of the type $e^+ \, e^- \, \bar{t} \, q$ (also electroweak operators) were discussed in~\cite{BarShalom:1999iy}. Finally, a complete set of relations for the operators involved in top-FCNC physics can be found in~\cite{AguilarSaavedra:2008zc}. All results discussed in the next sections take into account all the operators described in the above references relevant to top-FCNC physics at an electron-positron or a photon-photon collider.
\section{The story so far and the LHC}
\label{sec:story}

The search for new top FCNC physics started with indirect measurements of the branching ratios of top decaying to $qZ$, $q \gamma$ and $qg$ at LEP (q stands for the sum of $u$ and $c$-quarks). Indirect measurement are bounds on the branching ratios that have their origin in bounds on the cross sections of FCNC top production processes, with a subsequent decay $t \to b W$. This translation is correct if only one coupling constant describes the interaction $\bar{t} q V$, where $q=u,c$ and $V$ is a gauge boson. As a simple example, the production cross section for $e^+ e^- \to \bar{t} q$, has contribution from operators of the type $\bar{t} q V$ but also from four-fermion operators. Therefore, a measurement of this cross section will not allow, in the most general case, to put a bound on any of the branching ratios $BR (t \to q V)$. Moreover, bounds on cross sections that are converted on bounds on the branching ratios rely on the fact that the experimental analysis is not contaminated with other physical processes that would invalidate the conversion.
\begin{table}[h!]
\begin{center}
  \begin{tabular}{  l  c  c  c }
    \hline
     & LEP & HERA & Tevatron  \\ \hline \hline
    $Br(t \rightarrow q \, Z)$      & $ < \, 7.8 \% \,$ & $  < \, 49\% \,$
    & $ < \, 3.7 \% \,^d$  \\ \hline
    $Br(t \rightarrow q \, \gamma)$ & $ < \, 2.4 \% \,$ & $ < \, 0.64 \% (u) \,$
    & $ < \, 3.2 \% \,^d$  \\ \hline
    $Br(t \rightarrow q \, g)$      & $ < \, 17 \% \,$ & $ < \, 13 \% \,$
    & $ < \,  0.045 \% \,$ \\
    \hline
  \end{tabular}
\end{center}
\vskip -0.5cm
\caption{Current experimental bounds on FCNC branching ratios. The superscript "d" refers to bounds obtained from direct measurements, as explained in the text.} \label{tab:oldexplimits}
\end{table}
In table \ref{tab:oldexplimits} (see~\cite{Coimbra:2008qp} for a complete list of references) we present the experimental limits obtained at LEP, HERA and at the Tevatron. The superscript "d" refers to bounds obtained from direct measurements, that is, from $t \bar{t}$ production with one of the top-quarks decaying to $bW$ and the other to $qV$, with $q=u,c$ and $V=Z, \gamma$. The best bounds are now at the $\%$ level, except for the indirect bound on $t \to q g$ which is $0.045 \%$, from the measurement of the direct top production cross section at the Tevatron.

The gauge structure of the SM implies that a given dimension 6 operator with an impact on top interactions can also have a parallel effect on processes involving only bottom quarks. The most recent analysis for top-FCNC operators using all available B physics data was performed in~\cite{Fox:2007in} (see also~\cite{Grzadkowski:2008mf}).
The underlying SM gauge structure gives rise to a hierarchy of constraints: the gauge structure manifests more strongly in the operators denoted by $LL$ in~\cite{Fox:2007in} as these operators are built with only $SU(2)$ doublets. Operators $RR$, built with singlets alone, are obviously the least constrained as no relation exists between a $R$-top and a $R$-bottom.
\begin{table}[h]
\begin{center}
  \begin{tabular}{  l  c  c  c c c c c}
    \hline
    & ${\cal O}_{\phi}^{LL}$ & ${\cal O}_{tW\phi}^{RL}$ & ${\cal O}_{tB\phi}^{RL}$ &  ${\cal O}_{tW\phi}^{LR}$ & ${\cal O}_{tB\phi}^{LR}$ & ${\cal O}_{\phi_t}^{RR}$ \\ \hline \hline
    $Br(t \rightarrow c \, Z)     $  & $ {\cal O}(10 ^{-6})$ & $3.4 \times 10 ^{-5}$ & $8.4 \times 10 ^{-6}$ & $4.5 \times 10 ^{-3}$ & d & d \\ \hline
    $Br(t \rightarrow c \, \gamma)$  & $ -       $ & $1.8 \times 10 ^{-5}$           & $4.8 \times 10 ^{-5}$ & $2.3 \times 10 ^{-3}$ & d & d  \\ \hline
    $Br(t \rightarrow u \, Z)     $  & ${\cal O} (10^{-5}) $ & $4.1 \times 10 ^{-5}$ & $1.2 \times 10 ^{-4}$ & $3.2 \times 10 ^{-3}$ & d & d \\ \hline
    $Br(t \rightarrow u \, \gamma)$  & $ -       $ & $2.1 \times 10 ^{-5}$           & $6.7 \times 10 ^{-4}$ & $1.6 \times 10 ^{-3}$ & d & d \\ \hline \hline
  \end{tabular}
\end{center}
\vskip -0.5cm
\caption{Bounds from B-physics obtained in~\cite{Fox:2007in}.}
\label{tab:bphysics}
\end{table}
In Table \ref{tab:bphysics} we present the set of constraints on the branching ratios obtained in~\cite{Fox:2007in} when only one operator is taken at a time. Considering the prediction for the LHC with an integrated luminosity of $100 \, fb^{-1}$~\cite{toni, fla, CMS}, as shown in Table~\ref{tab:ATLASCMSlim}, it is clear that, in this approximation, operators of the type $LL$ are already constrained beyond the reach of the LHC.
\begin{table}[h]
\begin{center}
  \begin{tabular}{  l  c   c}
    \hline
     & ATLAS \& CMS (10 $fb^{-1}$) & ATLAS \& CMS ($100 fb^{-1}$)  \\ \hline \hline
    $Br(t \rightarrow q \, Z)$      & $  2.0 \times 10 ^{-4}$ & $  4.2 \times 10 ^{-5}$   \\ \hline
    $Br(t \rightarrow q \, \gamma)$ & $  3.6 \times 10 ^{-5}$ & $  1.0 \times 10 ^{-5}$  \\ \hline
    $Br(t \rightarrow q \, g)$ (ATLAS)     & $  1.3 \times 10 ^{-3}$ & $   4.2 \times 10 ^{-4}$                  \\
    \hline
  \end{tabular}
\end{center}
\vskip -0.5cm
\caption{Direct bounds based on the processe $pp \to t \bar{t} \to bW \, \bar{q} X$ at 95 \% CL.}
\label{tab:ATLASCMSlim}
\end{table}
This is true for operators of type $LL$, while limits on $LR$ and $RL$ operators are close to what is expected to be measured at the LHC. B factories and the Tevatron are still collecting data and therefore these constraints will be even stronger by the time the LHC starts to analyse data. As B physics only constraints operators with origin in the electroweak sector the best bound on $Br(t \rightarrow q \, g)$ is still the Tevatron indirect bound. Note that the prediction for the LHC at 14 $TeV$ for $Br(t \rightarrow q \, g)$ (and $100 \, fb^{-1}$) is similar to the indirect Tevatron bound. Again, this is because the Tevatron's is an indirect bound - a similar analysis for direct top production at the LHC for a 14 $TeV$ center-of-mass energy and $10 \, fb^{-1}$ integrated luminosity (ATLAS only) gives $Br(t \rightarrow q \, g) < 9 \times 10^{-5}$~\cite{chengdias}.

So far we have discussed bounds on the branching ratios even if some of them stem from limits on the productions cross sections, where FCNC is present, by taking one operator at the time. However, those bounds can also be used to place restrictions on the operators themselves. This is particularly true when four-fermion operators are present because those operators do not contribute to any of the top FCNC branching ratios discussed so far. In fact, four-fermion operators contribute only to $Br(t \rightarrow q \, e^+ e^-)$, a process that was not studied at the LHC. Both theoretical~\cite{BarShalom:1999iy} and experimental (LEP)~\cite{filipe} studies were performed for the four-fermion operators and restriction on the four-fermion coupling constants were set. It was shown in~\cite{BarShalom:1999iy} that the direct bounds on the four-fermion coupling constants will improve at a future electron-positron collider which is a consequence of the corresponding production cross sections growth with energy. Therefore, this is clearly a case where bounds on couplings will definitely improve with the next generation of electron-positron colliders.
\section{Is there top FCNC left to explore?}

Several studies dedicated to top production and decay involving FCNC couplings were performed for electron-positron colliders. Direct bounds based on the process $e^+ e^- \to t \bar{t} \to bW \, \bar{q} X$ were calculated in~\cite{AguilarSaavedra:2000db} for $\sqrt{s}=500 \, GeV$ and $\sqrt{s}=800 \, GeV$. In Table~\ref{tab:limitsjuan} we present limits for $Br(t \rightarrow q \, Z)$ and $Br(t \rightarrow q \, \gamma)$ taken from~\cite{AguilarSaavedra:2000db} for $\sqrt{s}=500 \, GeV$ and $300 fb^{-1}$ of integrated luminosity. The bounds degrade as the center-of-mass energy rises due to a decreasing $t \bar{t}$ production cross section and improve as the center-of-mass energy approaches the $t \bar{t}$ threshold.
\begin{wraptable}{l}{0.5\columnwidth}
\centerline{\begin{tabular}{  l  c }
    \hline
     & $\sqrt{s} = 500$ GeV $(300 fb^{-1})$   \\ \hline \hline
    $Br(t \rightarrow q \, Z)$      & $ {\cal O} (10^{-3})$      \\ \hline
    $Br(t \rightarrow q \, \gamma)$ & $ {\cal O} (10^{-4})$     \\ \hline \hline
  \end{tabular}}
\caption{Direct bounds based on the process $e^+ e^- \to t \bar{t} \to bW \, \bar{q} X$ at 95 \% CL.}
\label{tab:limitsjuan}
\end{wraptable}
There are several analysis of single top production where FCNC is present in the production process and not in the decay. Process $e^+ e^- \to t \bar{c} + c \bar{t}$, where $t \to b W$, was studied in~\cite{Han:1998yr} in an effective Lagrangian approach, using the most general top FCNC three-point interactions. The same process was discussed in the same approach but with the inclusion of the four-fermion operators in~\cite{BarShalom:1999iy}. Other top-FCNC production processes like $e^+ e^- \to t \bar{c}  \nu \bar{\nu} $ and $e^+ e^- \to t \bar{c} e^+ e^- $ were studied in~\cite{BarShalom:1997tm, BarShalom:1999iy}. We will use the results obtained for future electron-positron colliders to understand if these predictions can improve the bounds on all or some of the FCNC branching ratios after the LHC has collected 100 $fb^{-1}$ per experiment at $\sqrt{s}=14 \, TeV$.


To simplify our study we have considered all coupling constants real. We have made a further simplification by requiring that operators of type $O_{ij}$ cannot be distinguished from operators of type $O_{ji}$, that is, all operators are independent of where the top quark is placed. We have checked that these approximations do not affect our conclusions.
\begin{figure}[h]
\begin{center}
\includegraphics[width=7.2cm]{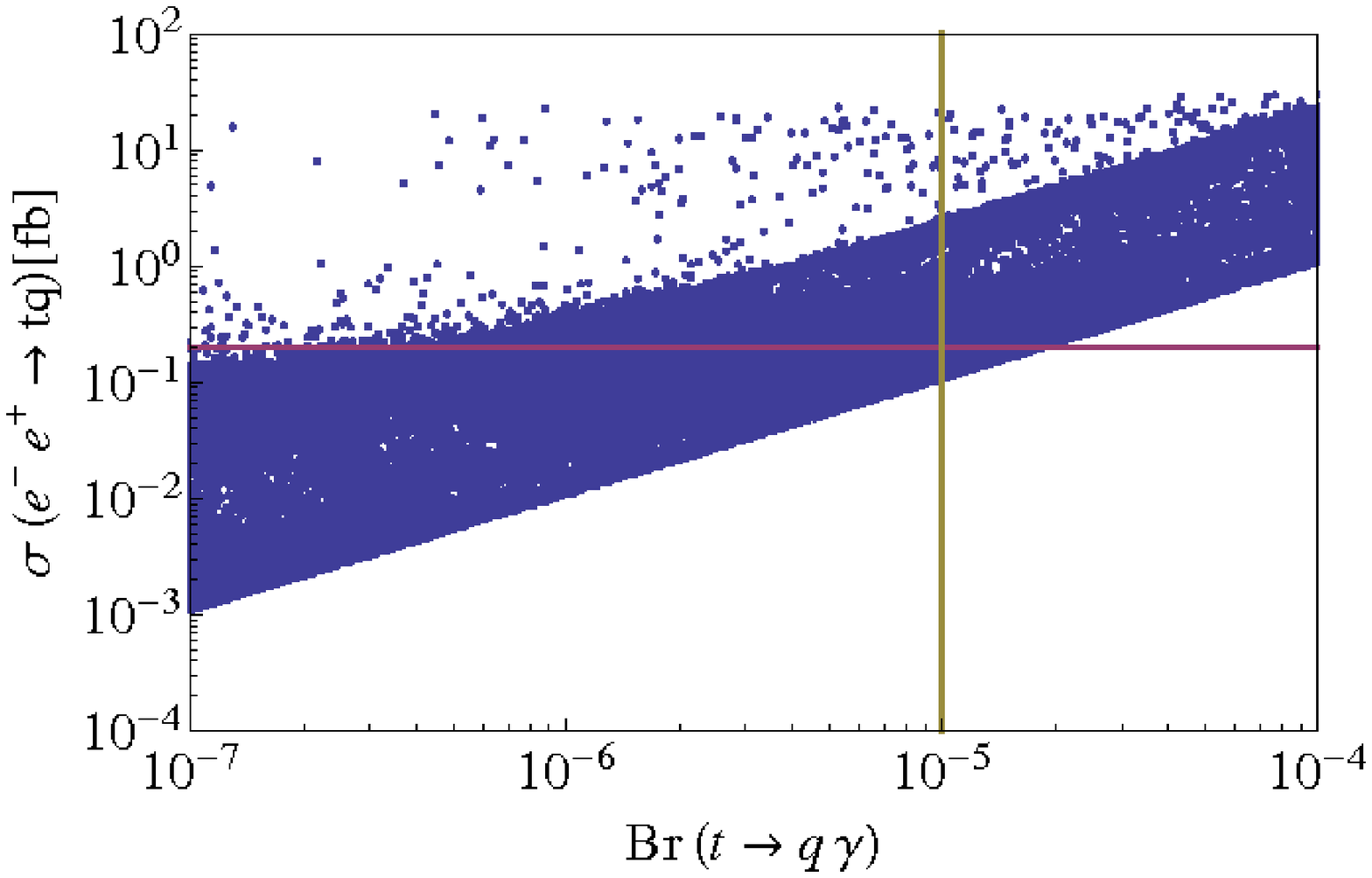}
\includegraphics[width=6.6cm]{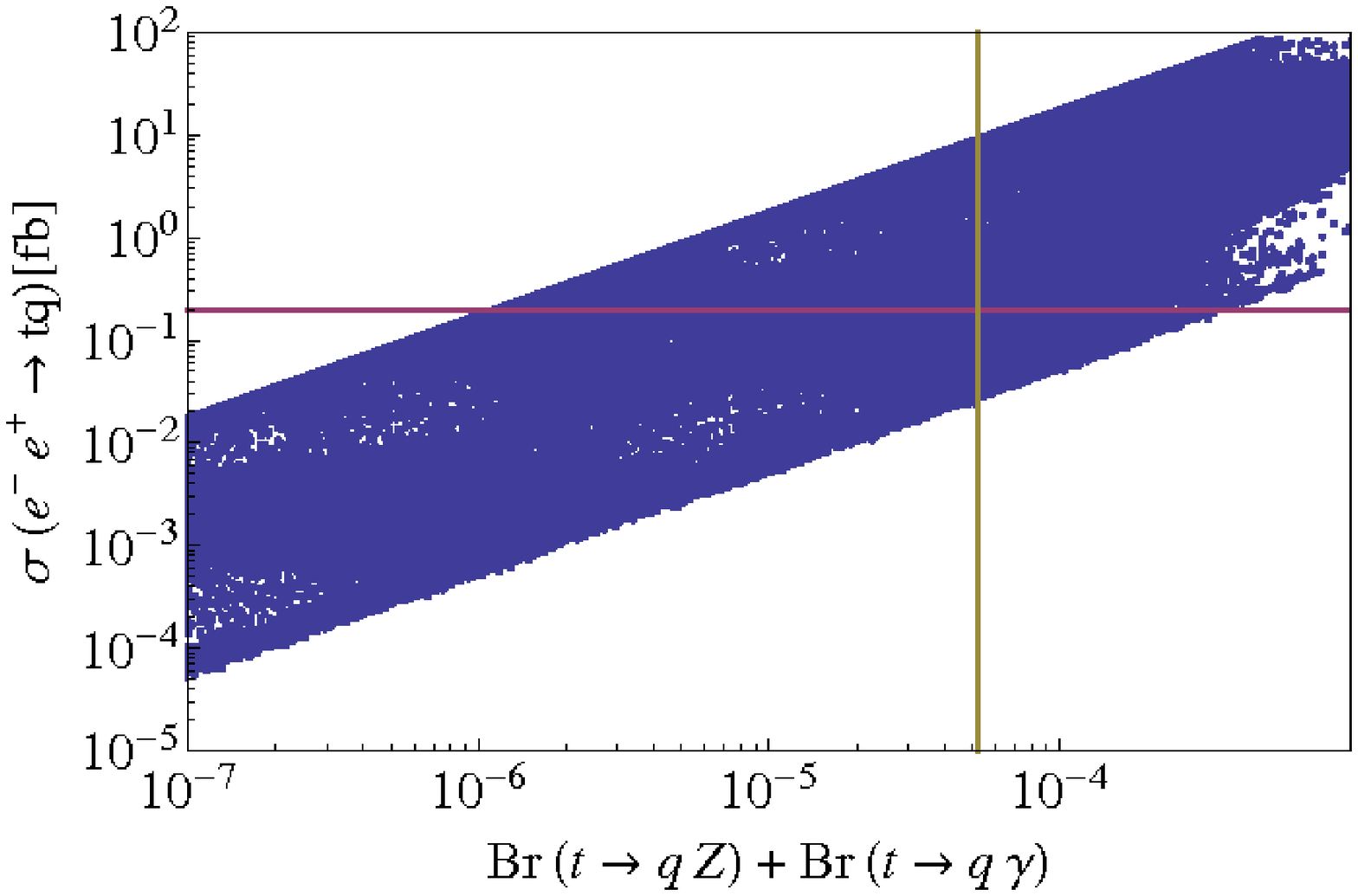}
\end{center}
\vspace{-0.6cm}
\caption{$\sigma_{e^+ e^- \to t \bar{c} +  c \bar{t}}$ as a function of the branching ratio $BR(t \to q \gamma)$ (left) and $BR(t \to q \gamma)+BR(t \to q Z)$ (right) with $q=u,c$ for $\sqrt{s}=500 \, GeV$. }
\label{fig:eetc}
\end{figure}
We have randomly generated 400K points for the coupling constants written as "$a \, 10^{b}$", with $-5 < a < 5$ and $-8  < b < -1$. In Figure~\ref{fig:eetc} we present the $e^+ e^- \to t \bar{c} +  c \bar{t}$ cross section, for $\sqrt{s} = 500 \, GeV$ as a function of the branching ratio $BR(t \to q \gamma)$ (left) and $BR(t \to q \gamma)+BR(t \to q Z)$ (right) with $q=u,c$.
\begin{figure}[h]
\begin{center}
\includegraphics[width=6.9cm]{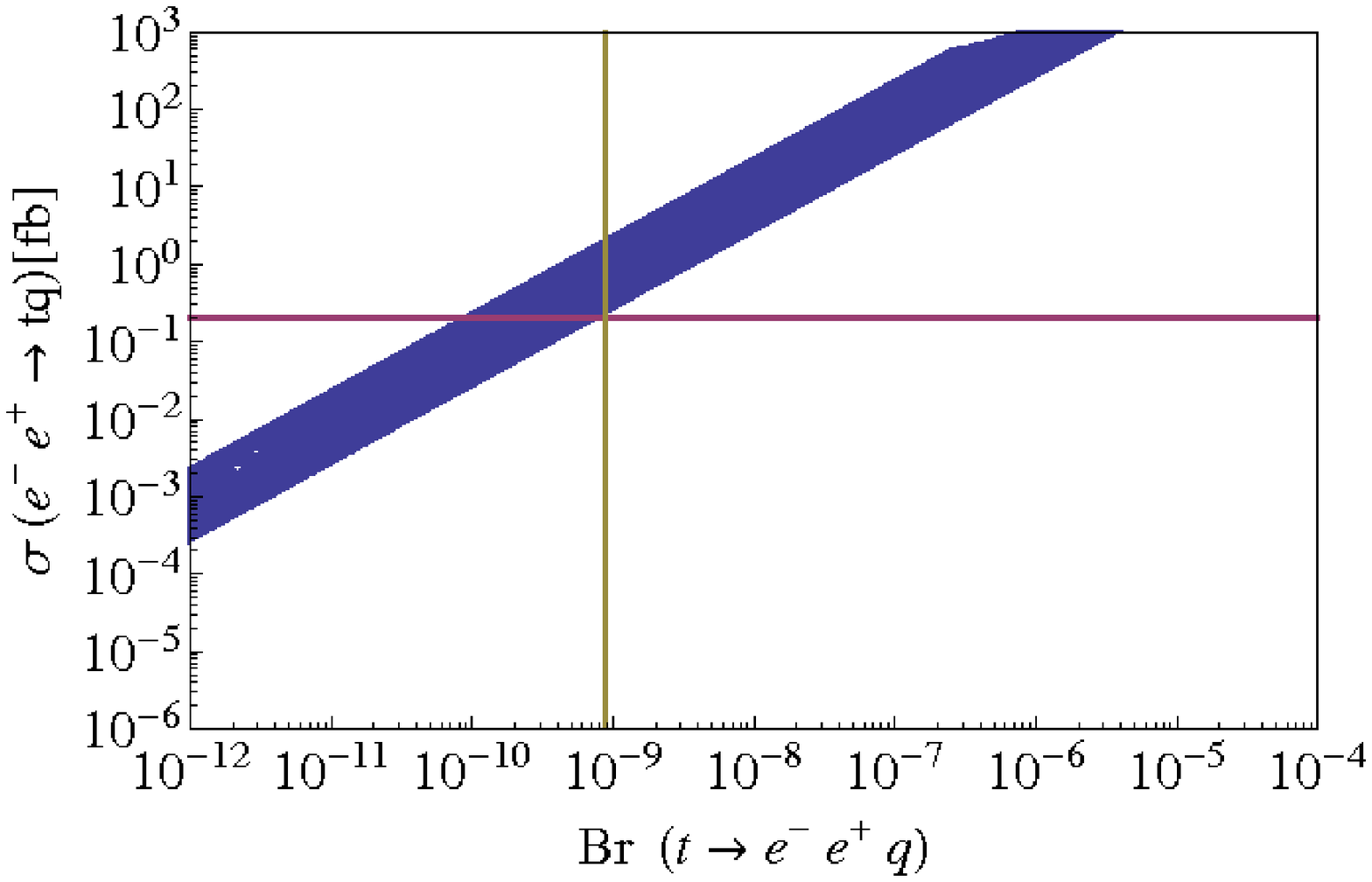}
\includegraphics[width=6.9cm]{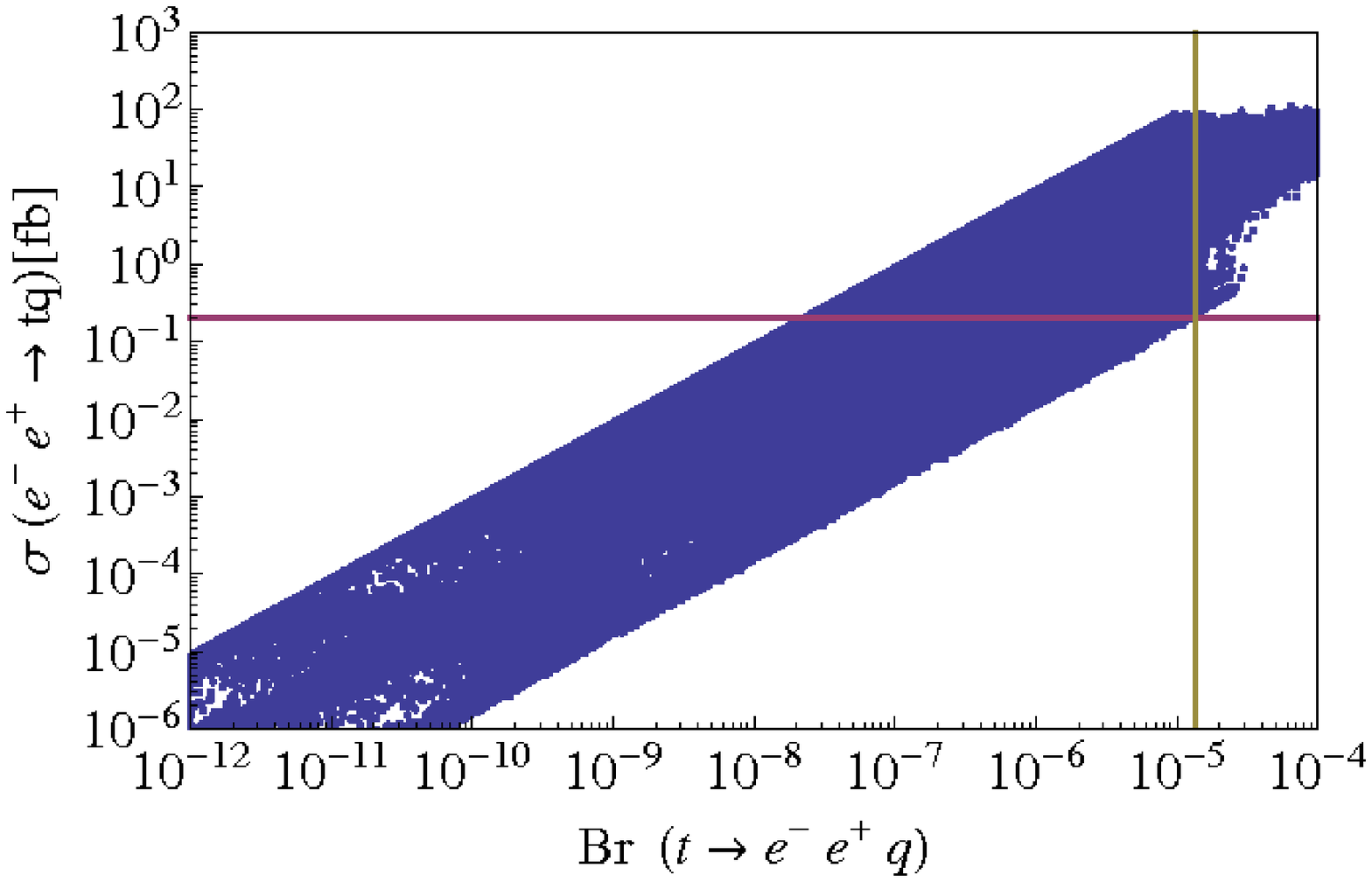}
\end{center}
\vspace{-0.6cm}
\caption{$\sigma_{e^+ e^- \to t \bar{q} + q \bar{t}}$ as a function of the branching ratio $BR(t \to q e^+ e^-)$  with $q=u,c$ with only four-fermion operators (left) and no four-fermion operators (right). }
\label{fig:eetc2}
\end{figure}
First we should note that when all couplings are taken into account, there is no simple proportionality between cross section and branching ratio. However a bound on the cross section can still be translated to a bound on a branching ratio. In the figures we draw a horizontal line that correspond to the upper limit set by the analysis in~\cite{Han:1998yr} for $\sqrt{s}=500 \, GeV$ and a luminosity of 500 $fb^{-1}$. The vertical line in Figure~\ref{fig:eetc} corresponds to the 14 $TeV$ LHC prediction for 100 $fb^{-1}$. We conclude that, because the lines cross inside the painted region, even if close to the border, the bound can only be improved with either an increase in luminosity or in center-of-mass of the electron-positron machine~\cite{Han:1998yr}.

In Figure \ref{fig:eetc2} we discuss how the inclusion of just one set of operators can affect the bounds on the branching ratios. There is still no bound or prediction available for $BR(t \to q e^+ e^-)$, but the LHC has the means to do it. It will probably be of the same order of magnitude of the one for $BR(t \to Z q)$, that is, $ {\cal O} (10^{-4})$. In the left panel we present the cross section for $e^+ e^- \to t \bar{q} + q \bar{t}$ as a function of $BR(t \to q e^+ e^-)$ with only four-fermion operators. The horizontal line is the same as the one in the previous plots while the vertical line points to a rough estimate of the bound that would be set on $BR(t \to q e^+ e^-)$. When only four-fermion operators are taken into account this bound is $ {\cal O} (10^{-9})$. In the right panel we present the same plot but without the four-fermion operators. In this case the bound is $ {\cal O} (10^{-5})$. If all operators are taken into account the bound becomes unreliable due to interference terms but if any it will always be worst than $ {\cal O} (10^{-5})$.

The simplest process that could probe the strong FCNC branching ratio is $e^+ e^- \to t \bar{q} g$. In this case FCNC could come either from the strong, from the electroweak, or from the four-fermion sector. We have checked that the bound is several orders of magnitude worst than what is expected for the equivalent indirect bound at the LHC.


\begin{wrapfigure}{l}{0.6\textwidth}
  \begin{center}
    \includegraphics[width=7.cm]{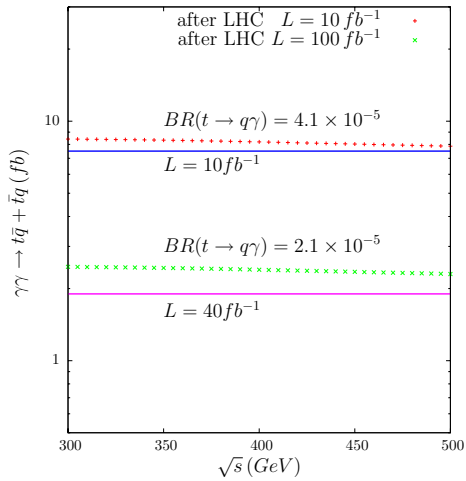}
  \end{center}
  \vskip -0.5cm
  \caption{$\sigma_{\gamma \gamma \to t \bar{c} + c \bar{t}}$ as a function of $\sqrt{s}$. We present the LHC bounds for $BR(t \to q \gamma)$ for integrated luminosities of 10 and 100 $fb^{-1}$ }
\label{fig:gamma}
\end{wrapfigure}
Finally we will just briefly comment on the bounds that can be obtained for a $\gamma \gamma$ collider. This is a particularly interesting case because, in our framework,  there is only one coupling constant involved in the process $\gamma \gamma \to t \bar{q} + q \bar{t}$ which means that it can be unambiguous translated to a limit on $BR (t \to \bar{q} \gamma)$.
%
A detailed study was presented in~\cite{Abraham:1997zf} (see also \cite{Boos:2000gr}) for $\sqrt{s} = 400, \, 500, \,$ and $800 \, GeV$ center-of-mass energies. Using their results and the predicted bounds for $BR(t \to q \gamma)$ with $q=u,c$ for the 14 $TeV$ LHC for integrated luminosities of 10 and 100 $fb^{-1}$, it is clear that the LHC bound on $BR(t \to q \gamma)$ has a good chance to be improved at a future $\gamma \gamma$ collider. Figure~\ref{fig:gamma} shows that a total luminosity of 40 $fb^{-1}$ would be enough to overcame the LHC 100  $fb^{-1}$  bound almost independently of the $\gamma \gamma$ collider center-of mass energy.
\section{Conclusions}

In this work we have discussed what will be left to study in top FCNC physics after the LHC has reached a stage where no significant change in FCNC bounds will occur. We have concentrated on the scenario where no evidence for new physics is found - otherwise, the role of a linear collider has still to be investigated. Taking into account the predictions done so far for an electron-positron ($\gamma \gamma$) collider our conclusions are as follows
\begin{itemize}

\item In this scenario, improving the LHC bounds on the BR, depends on the energy and especially on the luminosity of the future collider. Taking as a benchmark the available studies no significant improvement on the bounds of the branching ratios is expected. If new physics is founds, particular operators can be probed with definite observables.

\item Regarding the four-fermion operators, the bounds on the coupling constants will certainly improve due to rise of cross section with the collider's energy.

\item Improvement on other specific couplings taken one at a time can also be achieved. We did not consider those scenarios in our study.

\item A photon-photon collider will most certainly improve the bound on the $t \to q \gamma$ FCNC branching ratio.

\item Finally, NLO QCD corrections to top FNCN decays to $Z$ and $\gamma$ were shown to be negligible for our choice of operators~\cite{Zhang:2008yn}. Correction to $t \to q g$, $q=u,c$ were shown to be of the order of 20 \%~\cite{Zhang:2008yn}.

\end{itemize}

\section{Acknowledgments}

We thank Filipe Veloso and Ant\'onio Onofre for discussions.


\begin{footnotesize}


\end{footnotesize}



\begin{thebibliography}{99}


\bibitem{Eilam:1990zc}
  G.~Eilam, J.~L.~Hewett and A.~Soni,
  Phys.\ Rev.\  D {\bf 44} (1991) 1473
  [Erratum-ibid.\  D {\bf 59} (1999) 039901].

\bibitem{AguilarSaavedra:2004wm}
  J.~A.~Aguilar-Saavedra,
  Acta Phys.\ Polon.\  B {\bf 35} (2004) 2695.

\bibitem{AguilarSaavedra:2002ns}
  J.~A.~Aguilar-Saavedra and B.~M.~Nobre,
  Phys.\ Lett.\  B {\bf 553} (2003) 251.

\bibitem{Atwood:1996vj}
  D.~Atwood, L.~Reina and A.~Soni,
  Phys.\ Rev.\  D {\bf 55} (1997) 3156.


\bibitem{He:1998ie}
  H.~J.~He and C.~P.~Yuan,
  Phys.\ Rev.\ Lett.\  {\bf 83} (1999) 28.




\bibitem{Buchmuller:1985jz}
  W.~Buchmuller and D.~Wyler,
  Nucl.\ Phys.\  B {\bf 268} (1986) 621.

\bibitem{Ferreira:2005dr}
  P.~M.~Ferreira, O.~Oliveira and R.~Santos,
  Phys.\ Rev.\  D {\bf 73} (2006) 034011;
  P.~M.~Ferreira and R.~Santos,
  Phys.\ Rev.\  D {\bf 73} (2006) 054025.

\bibitem{Han:1998yr}
  T.~Han and J.~L.~Hewett,
  Phys.\ Rev.\  D {\bf 60} (1999) 074015.

\bibitem{Grzadkowski:2003tf}
  B.~Grzadkowski, Z.~Hioki, K.~Ohkuma and J.~Wudka,
  Nucl.\ Phys.\  B {\bf 689} (2004) 108.

\bibitem{Ferreira:2008cj}
  P.~M.~Ferreira, R.~B.~Guedes and R.~Santos,
  Phys.\ Rev.\  D {\bf 77} (2008) 114008.


\bibitem{Coimbra:2008qp}
  R.~A.~Coimbra, P.~M.~Ferreira, R.~B.~Guedes, O.~Oliveira, A.~Onofre, R.~Santos and M.~Won,
  Phys.\ Rev.\  D {\bf 79} (2009) 014006.





\bibitem{BarShalom:1999iy}
  S.~Bar-Shalom and J.~Wudka,
  Phys.\ Rev.\  D {\bf 60} (1999) 094016.

\bibitem{AguilarSaavedra:2008zc}
  J.~A.~Aguilar-Saavedra,
  Nucl.\ Phys.\  B {\bf 812} (2009) 181.


\bibitem{Fox:2007in}
  P.~J.~Fox, Z.~Ligeti, M.~Papucci, G.~Perez and M.~D.~Schwartz,
  Phys.\ Rev.\  D {\bf 78} (2008) 054008.

\bibitem{Grzadkowski:2008mf}
  B.~Grzadkowski and M.~Misiak,
  Phys.\ Rev.\  D {\bf 78} (2008) 077501;
  Y.~Grossman, Z.~Ligeti and Y.~Nir,
  arXiv:0904.4262 [hep-ph].

\bibitem{toni} J.~Carvalho {\emph et al.}, {\em~Eur.~Phys.~J.}{\bf C 52} (2007)
999-1019.

\bibitem{fla} T. Lari {\em et al}, Report of Working Group 1 of the CERN Workshop
``Flavour in the era of the LHC'', hep-ph/0801.1800.

\bibitem{CMS} CMS Physics TDR: Volume II, CERN/LHCC 2006-021, http://cmsdoc.cern.ch/cms/cpt/tdr/.

\bibitem{chengdias} T.~L.~Cheng and P.~Teixeira-Dias, ATL-PHYS-PUB-2006-029.


\bibitem{filipe} DELPHI collaboration, S.~Andringa \textit{et al}, DELPHI 2006-003 CONF 749, 2006. F.~Veloso, Master Thesis, Univ. T\'ecnica de Lisboa, 2004.


\bibitem{AguilarSaavedra:2000db}
  J.~A.~Aguilar-Saavedra,
  Phys.\ Lett.\  B {\bf 502} (2001) 115;
  J.~A.~Aguilar-Saavedra and T.~Riemann,
  arXiv:hep-ph/0102197.

\bibitem{BarShalom:1997tm}
  S.~Bar-Shalom, G.~Eilam, A.~Soni and J.~Wudka,
  Phys.\ Rev.\ Lett.\  {\bf 79} (1997) 1217.

\bibitem{Abraham:1997zf}
  K.~J.~Abraham, K.~Whisnant and B.~L.~Young,
  Phys.\ Lett.\  B {\bf 419} (1998) 381.

\bibitem{Boos:2000gr}
  E.~E.~Boos,
  Nucl.\ Instrum.\ Meth.\  A {\bf 472} (2001) 22.

\bibitem{Zhang:2008yn}
  J.~J.~Zhang, C.~S.~Li, J.~Gao, H.~Zhang, Z.~Li, C.~P.~Yuan and T.~C.~Yuan,
  Phys.\ Rev.\ Lett.\  {\bf 102} (2009) 072001.


\end{thebibliography}
\end{document}